\documentclass[prd,superscriptaddress,twocolumn,nofootinbib,showpacs,preprintnumbers,floatfix]{revtex4}

\usepackage{amsfonts}
\usepackage{mathrsfs}
\usepackage{amssymb}
\usepackage{amsmath}
\usepackage{graphicx}
\usepackage{epsfig}
\usepackage{verbatim}

\begin{document}

\setlength{\parskip}{0cm}

\title{ Blueprints of the No-Scale Multiverse at the LHC }

\author{Tianjun Li}

\affiliation{Key Laboratory of Frontiers in Theoretical Physics, Institute of Theoretical Physics, Chinese Academy of Sciences, Beijing 100190, P. R. China }

\affiliation{George P. and Cynthia W. Mitchell Institute for Fundamental Physics and Astronomy, Texas A$\&$M University, College Station, TX 77843, USA }

\author{James A. Maxin}

\affiliation{George P. and Cynthia W. Mitchell Institute for Fundamental Physics and Astronomy, Texas A$\&$M University, College Station, TX 77843, USA }

\author{Dimitri V. Nanopoulos}

\affiliation{George P. and Cynthia W. Mitchell Institute for Fundamental Physics and Astronomy, Texas A$\&$M University, College Station, TX 77843, USA }

\affiliation{Astroparticle Physics Group, Houston Advanced Research Center (HARC), Mitchell Campus, Woodlands, TX 77381, USA}

\affiliation{Academy of Athens, Division of Natural Sciences, 28 Panepistimiou Avenue, Athens 10679, Greece } 

\author{Joel W. Walker}

\affiliation{Department of Physics, Sam Houston State University, Huntsville, TX 77341, USA }

%%%%%%%%%%%%%%%%%%%%%%%%%%%%%%%%%%%%%%%%%%%%%%%%%%%%%%%%%%%%%%%%%%%%%%%%%%%%

\begin{abstract}

We present a contemporary perspective on the String Landscape and the Multiverse of
plausible string, M- and F-theory vacua.  In contrast to traditional statistical
classifications and capitulation to the anthropic principle, we seek only to demonstrate
the existence of a non-zero probability for a universe matching
our own observed physics within the solution ensemble.  We argue for the importance of No-Scale
Supergravity as an essential common underpinning for the spontaneous emergence of a cosmologically
flat universe from the quantum ``nothingness''.  Concretely, we continue to probe the phenomenology of
a specific model which is testable at the LHC and Tevatron.  Dubbed No-Scale ${\cal F}$-$SU(5)$,
it represents the intersection of the Flipped $SU(5)$ Grand Unified Theory (GUT)
with extra TeV-Scale vector-like multiplets derived
out of F-theory, and the dynamics of No-Scale Supergravity, which in turn imply
a very restricted set of high energy boundary conditions.
By secondarily minimizing the minimum of the scalar Higgs potential, we 
dynamically determine the ratio $\tan \beta \simeq 15-20$ of up- to down-type Higgs vacuum
expectation values (VEVs), the universal gaugino boundary mass $M_{1/2} \simeq 450$~GeV,
and consequently also the total magnitude of the GUT-scale Higgs VEVs,
while constraining the low energy Standard Model gauge couplings.
In particular, this local {\it minimum minimorum} lies within the previously described
``golden strip'', satisfying all current experimental constraints.  We emphasize, however,
that the overarching goal is not to establish why our own particular universe possesses
any number of specific characteristics, but rather to tease out what generic principles
might govern the superset of all possible universes.

\end{abstract}


\preprint{ACT-01-11, MIFPA-11-03}

\maketitle

%%%%%%%%%%%%%%%%%%%%%%%%%%%%%%%%%%%%%%%%%%%%%%%%%%%%%%%%%%%%%%%%%%%%%%%%%%%%%

\section{Introduction}

The number of consistent, meta-stable vacua of string, M- or (predominantly)
F-theory flux compactifications which exhibit broadly plausible phenomenology,
including moduli stabilization and broken
supersymmetry~\cite{Bousso:2000xa, Giddings:2001yu, Kachru:2003aw, Susskind:2003kw, Denef:2004ze, Denef:2004cf}, is popularly
estimated~\cite{Denef:2004dm,Denef:2007pq} to be of order $10^{500}$.  It is moreover
currently in vogue to suggest that degeneracy of common features
across these many ``universes'' might statistically isolate the physically realistic
universe from the vast ``landscape'', much as the entropy function coaxes the singular order
of macroscopic thermodynamics from the chaotic duplicity of the entangled quantum microstate.
We argue here though the counter point that we are not obliged {\it a priori} to live
in the likeliest of all universes, but only in one which is possible.  The existence
merely of a non-zero probability for our existence is sufficient.

We indulge for this effort the fanciful imagination that the ``Multiverse'' of string
vacua might exhibit some literal realization beyond our own physical sphere.
A single electron may be said to wander all histories through interfering apertures,
though its arrival is ultimately registered at a localized point on the target.
The journey to that destination is steered by the full dynamics of the theory, although
the isolated spontaneous solution reflects only faintly the richness of the solution ensemble.
Whether the Multiverse be reverie or reality, the conceptual superset of our own physics
which it embodies must certainly represent the interference of all navigable universal histories.

Surely many times afore has mankind's notion of the heavens expanded - the Earth dispatched
from its central pedestal in our solar system and the Sun rendered one among some hundred billion
stars of the Milky Way, itself reduced to one among some hundred billion galaxies.
%Modern cosmology indeed suggests that not only are we not at the center of our universe, there
%simply is no center.  The human capacity for awe is dealt equal exhaustion from the other
%end of the scale, the Planck length $\mathcal{O}\,(10^{-35})$~m being comparatively smaller
%than our bodies by several orders of magnitude more than the visible light horizon is larger.
Finally perhaps, we come to the completion of our Odyssey, by realizing that our
Universe is one of at least $10^{500}$ so possible, thus rendering the anthropic view
of our position in the Universe (environmental coincidences explained away by the availability of
$10^{11} \times 10^{11}$ solar systems) functionally equivalent to the anthropic view of the
origin of the Universe (coincidences in the form and content of physical laws explained away
by the availability, through dynamical phase transitions, of $10^{500}$ universes).
Nature's bounty has anyway invariably trumped our wildest anticipations, and though
frugal and equanimous in law, she has spared no extravagance or whimsy in its manifestation.

Our perspective should not be misconstrued, however, as complacent retreat into the tautology of
the weak anthropic principle.  It is indeed unassailable truism that an observed universe
must afford and sustain the life of the observer, including requisite constraints,
for example, on the cosmological constant~\cite{Weinberg:1987dv} and gauge hierarchy.
Our point of view, though, is sharply different; we should be able to
resolve the cosmological constant and gauge hierarchy problems through investigation
of the fundamental laws of our (or any single) Universe, its accidental and specific properties
notwithstanding, without resorting to the existence of observers.  In our view, the observer is the
output of, not the {\it raison d'\^etre} of, our Universe.  Thus, our attention is advance from this
base camp of our own physics, as unlikely an appointment as it may be, to the summit goal of the
master theory and symmetries which govern all possible universes.  In so seeking, our first
halting forage must be that of a concrete string model which can describe Nature locally.

%%%%%%%%%%%%%%%%%%%%%%%%%%%%%%%%%%%%%%%%%%%%%%%%%%%%%%%%%%%%%%%%%%%%%%%%%%%%%

\section{The Ensemble Multiverse}

The greatest mystery of Nature is the origin of the Universe itself.
Modern cosmology is relatively clear regarding the occurrence of a hot big bang,
and subsequent Planck, grand unification, cosmic inflation, lepto- and baryogenesis,
and electroweak epochs, followed by nucleosynthesis, radiation decoupling, and large
scale structure formation.  In particular, cosmic inflation can address the flatness and
monopole problems, explain homogeneity, and generate the fractional anisotropy of the
cosmic background radiation by quantum fluctuation of the inflaton
field~\cite{Guth:1980zm, Linde:1981mu, Albrecht:1982wi, Ellis:1982ws, Nanopoulos:1982bv}.
A key question though, is from whence the energy of the Universe arose.  Interestingly,
the gravitational field in an inflationary scenario can supply the required positive
mass-kinetic energy, since its potential energy becomes negative without bound,
allowing that the total energy could be exactly zero.

Perhaps the most striking revelation of the post-WMAP~\cite{Spergel:2003cb, Spergel:2006hy, Komatsu:2010fb} era is the decisive
determination that our Universe is indeed globally flat, {\it i.e.}~with the net energy contributions from
baryonic matter $\simeq 5\%$, dark matter $\simeq 23\%$, and the cosmological constant (dark energy)
$\simeq 72\%$ finely balanced against the gravitational potential.  Not long ago, it was
possible to imagine the Universe, with all of its physics intact, hosting any arbitrary
mass-energy density, such that ``$k=+1$'' would represent a super-critical cosmology of
positive curvature, and ``$k=-1$'' the sub-critical case of negative curvature.  In
hindsight, this may come to seem as na\"{\i}ve as the notion of an empty infinite
Cartesian space.  The observed energy balance is highly suggestive of a fundamental symmetry
which protects the ``$k=0$'' critical solution, such that the physical constants of our
Universe may not be divorced from its net content.

This null energy condition licenses the speculative connection {\it ex nihilo} of our present
universe back to the primordial quantum fluctuation of an external system. Indeed, there is
nothing which quantum mechanics abhors more than nothingness.  This being the case, an extra
universe here or there might rightly be considered no extra trouble at all!  Specifically, it has
been suggested~\cite{Guth:1980zm, Linde:1981mu, Albrecht:1982wi, Steinhardt, Vilenkin:1983xq}
that the fluctuations of a dynamically evolved expanding universe might spontaneously produce
tunneling from a false vacuum into an adjacent (likely also false) meta-stable vacuum of lower
energy, driving a local inflationary phase, much as a crystal of ice or a bubble of steam may nucleate
and expand in a super-cooled or super-heated fluid during first order transition.  In this ``eternal
inflation'' scenario, such patches of space will volumetrically dominate by virtue of their
exponential expansion, recursively generating an infinite fractal array of causally disconnected
``Russian doll'' universes, nesting each within another, and each featuring
its own unique physical parameters and physical laws.
%inflation onsets at GUT scale, dual scales of F-SU5

From just the specific location on the solution ``target'' where our own Universe landed, it may
be impossible to directly reconstruct the full theory.  Fundamentally, it may be impossible even in
principle to specify why our particular Universe is precisely as it is.  However, superstring theory and
its generalizations may yet present to us a loftier prize - the theory of the ensemble Multiverse.

%%%%%%%%%%%%%%%%%%%%%%%%%%%%%%%%%%%%%%%%%%%%%%%%%%%%%%%%%%%%%%%%%%%%%%%%%%%%%

\section{The Invariance of Flatness}

More important than any differences between various possible vacua are the
properties which might be invariant, protected by basic symmetries of the underlying mechanics.
We suppose that one such basic property must be cosmological flatness, so that
the seedling universe may transition dynamically across the boundary of its own creation,
maintaining a zero balance of some suitably defined energy function.  In practice, this
implies that gravity must be ubiquitous, its negative potential energy allowing for positive
mass and kinetic energy.  Within such a universe, quantum fluctuations may not again cause
isolated material objects to spring into existence, as their net energy must necessarily be
positive.  For the example of a particle with 
mass $m$ on the surface of the Earth, the ratio of gravitational
to mass energy is more than nine orders of magnitude too small
\begin{equation}
\left| -\frac{G_N M_E m}{R_E}  \right| \div  m c^2 \simeq 7 \times 10^{-10}~,~
\end{equation}
where $G_N$ is the gravitational constant, $c$ is the speed of light, 
and $M_E$ and $R_E$ are the mass and radius of the Earth, respectively.
Even in the limiting case of a Schwarzschild black hole of mass $M_{BH}$, 
a particle of mass $m$ at the horizon $R_{S}=2 G_N  M_{BH}/c^2$ has a gravitational potential 
which is only half of that required.
\begin{equation}
\left| -\frac{G_N M_{BH} m}{R_{S}} \right| = \frac{1}{2} m c^2
\end{equation}
It is important to note that while the energy density for the gravitational field is
surely negative in Newtonian mechanics, the global gravitational field energy is not
well defined in general relativity.  Unique prescriptions
for a stress-energy-momentum pseudotensor can be formulated though,
notably that of Landau and Lifshitz.  Any such stress-energy can,  however,
be made to vanish locally by general coordinate transformation, and it is 
not even entirely clear that the pseudotensor so applied is an appropriate general
relativistic object.  Given though that Newtonian gravity is the classical limit of 
general relativity, it is reasonable to suspect that the properly defined field
energy density will be likewise also negative, and that inflation is indeed consistent
with a correctly generalized notion of constant, zero total energy.

A universe would then be in this sense closed, an island unto itself, from the moment of
its inception from the quantum froth; only a universe {\it in toto} might so
originate, emerging as a critically bound structure possessing profound density and
minute proportion, each as accorded against intrinsically defined scales (the analogous
Newton and Planck parameters and the propagation speed of massless fields),
and expanding or inflating henceforth and eternally.

%%%%%%%%%%%%%%%%%%%%%%%%%%%%%%%%%%%%%%%%%%%%%%%%%%%%%%%%%%%%%%%%%%%%%%%%%%%%%

\section{The Invariance of No-Scale SUGRA}

Inflation, driven by the scalar inflaton field is itself inherently a quantum field
theoretic subject.  However, there is tension between quantum mechanics and general
relativity.  Currently, superstring theory is the best candidate for quantum gravity.  The
five consistent ten dimensional superstring theories, namely heterotic $E_8 \times E_8$,
heterotic $SO(32)$, Type I, Type IIA, Type IIB, can be unified by various duality transformations
under an eleven-dimensional M-theory~\cite{Witten:1995ex}, and the twelve-dimensional 
F-theory can be considered as the strongly coupled formulation of the Type IIB
string theory with a varying axion-dilaton field~\cite{Vafa:1996xn}.  Self consistency of the string
(or M-, F-) algebra implies a ten (or eleven, twelve) dimensional master spacetime,
some elements of which -- six (or seven, eight) to match our observed four large dimensions --
may be compactified on a manifold (typically Calabi-Yau manifolds
or $G_2$ manifolds) which conserves a requisite portion
of supersymmetric charges.

The structure of the curvature within the extra dimensions dictates in no small measure
the particular phenomenology of the unfolded dimensions, secreting away the ``closet space''
to encode the symmetries of all gauged interactions.  The physical volume of the internal 
spatial manifold is directly related to the effective Planck scale and basic gauge coupling
strengths in the external space.  The compactification is in turn described by fundamental
moduli fields which must be stabilized, {\it i.e.}~given suitable vacuum expectation values (VEVs).
The famous example of Kaluza and Klein prototypes the manner in which general covariance in five dimensions
is transformed to gravity plus Maxwell theory in four dimensions when the transverse fifth dimension
is cycled around a circle.  The connection of geometry to particle physics is perhaps nowhere
more intuitively clear than in the context of model building with $D6$-branes, where the
gauge structure and family replication are related directly to the brane stacking and intersection
multiplicities.  The Yukawa couplings and Higgs structure are in like manners also specified,
leading after radiative symmetry breaking of the chiral gauge sector to low energy masses
for the chiral fermions and broken gauge generators, each massless in the symmetric limit. 

From a top-down view, Supergravity (SUGRA) is an ubiquitous infrared limit of string theory,
and forms the starting point of any two-dimensional world sheet or D-dimensional target space action.
The mandatory localization of the Supersymmetry (SUSY) algebra, and thus the
momentum-energy (space-time translation) operators, leads to general coordinate
invariance of the action and an Einstein field theory limit.  Any available
flavor of Supergravity will not however suffice.  In general, extraneous
fine tuning is required to avoid a cosmological constant which scales like a dimensionally
suitable power of the Planck mass.  Neglecting even the question of whether such a universe might
be permitted to appear spontaneously, it would then be doomed to curl upon itself and collapse
within the order of the Planck time, for comparison about $10^{-43}$ seconds in our Universe.
Expansion and inflation appear to uniquely require properties which arise naturally only in the
No-Scale SUGRA formulation~\cite{Cremmer:1983bf,Ellis:1983sf, Ellis:1983ei, Ellis:1984bm, Lahanas:1986uc}.

SUSY is in this case broken while the vacuum energy density vanishes automatically at tree level due to
a suitable choice of the K\"ahler potential, the function which specifies the metric on superspace.
At the minimum of the null scalar potential, there are flat directions which leave the compactification
moduli VEVs undetermined by the classical equations of motion.  We thus receive without additional effort
an answer to the deep question of how these moduli are stabilized; they have been transformed into
dynamical variables which are to be determined by minimizing corrections to the scalar
potential at loop order.  In particular, the high energy gravitino mass $M_{3/2}$, and also the
proportionally equivalent universal gaugino mass $M_{1/2}$, will be established in this way.  Subsequently,
all gauge mediated SUSY breaking soft-terms will be dynamically evolved down from this boundary under the
renormalization group~\cite{Giudice:1998bp}, establishing in large measure the low energy phenomenology, and
solving also the Flavour Changing Neutral Current (FCNC) problem.  Since the moduli are fixed at a false local
minimum, phase transitions by quantum tunneling will naturally occur between discrete vacua.

We conjecture, for the reasons given prior, that the No-Scale SUGRA construction
could pervade all universes in the String Landscape with reasonable flux vacua.  This being the case,
intelligent creatures elsewhere in the Multiverse, though separated from us by a bridge too far,
might reasonably so concur after parallel examination of their own physics.  Moreover, they might leverage via
this insight a deeper knowledge of the underlying Multiverse-invariant master theory, of which our known string,
M-, and F-theories may compose some coherently overlapping patch of the garment edge.  Perhaps we yet share
appreciation, across the cords which bind our 13.7 billion years to their corresponding blink of history,
for the common timeless principles under which we are but two isolated condensations upon two particular
vacuum solutions among the physical ensemble.

%%%%%%%%%%%%%%%%%%%%%%%%%%%%%%%%%%%%%%%%%%%%%%%%%%%%%%%%%%%%%%%%%%%%%%%%%%%%%

\section{An Archetype Model Universe}

Though we engage in this work lofty and speculative questions of natural philosophy,
we balance abstraction against the measured material underpinnings of concrete
phenomenological models with direct and specific connection to tested and testable particle physics.
If the suggestion is correct that eternal inflation and No-Scale SUGRA models with
string origins together describe what is in fact our Multiverse, then we must as a prerequisite
settle the issue of whether our own phenomenology can be produced out of such a construction.

In the context of Type II intersecting D-brane models, we have indeed found one realistic 
Pati-Salam model which might describe Nature as we observe
it~\cite{Cvetic:2004ui, Chen:2007px, Chen:2007zu}. 
If only the F-terms of three complex structure moduli are non-zero, we also
automatically have vanishing vacuum energy, and obtain a generalized No-Scale SUGRA. 
It seems to us that the string derived Grand Unified Theories (GUTs), and particularly the 
Flipped $SU(5)\times U(1)_X$ models~\cite{Barr:1981qv, Derendinger:1983aj, Antoniadis:1987dx}, are also candidate
realistic string models with promising predictions that can be tested at the Large
Hadron Collider (LHC), the Tevatron, and other future experiments.

In the latter case, the Flipped $SU(5)\times U(1)_X$ gauge symmetry can be broken
down to the SM gauge symmetry by giving VEVs to  
one pair of the Higgs fields $H$ and $\overline{H}$ with quantum numbers
$(\mathbf{10}, \mathbf{1})$ and $(\mathbf{\overline{10}}, \mathbf{-1})$,
respectively.  The doublet-triplet 
splitting problem can be solved naturally
via the missing partner mechanism~\cite{Antoniadis:1987dx}.
Historically, Flipped $SU(5)\times U(1)_X$ models have been
constructed systematically in the free fermionic string 
constructions at Kac-Moody level one~\cite{Antoniadis:1987dx,Antoniadis:1987tv,Antoniadis:1988tt,Antoniadis:1989zy,Lopez:1992kg}.
To address the little hierarchy problem between the unification scale
and the string scale, the Testable Flipped $SU(5)\times U(1)_X$ model class was
proposed, which introduces extra TeV-scale vector-like particles~\cite{Jiang:2006hf}.
Models of this type have been constructed locally as examples of 
F-theory model building~\cite{Beasley:2008dc, Beasley:2008kw, Donagi:2008ca, Donagi:2008kj,Jiang:2009zza,Jiang:2009za}, and dubbed
${\cal F}$-$SU(5)$~\cite{Jiang:2009zza,Jiang:2009za} within that context.

Most recently, we have studied No-Scale extensions of the prior in
detail~\cite{Li:2010ws,Li:2010mi,Li:2010uu}, emphasizing the essential role of the
tripodal foundation formed by the $\mathcal{F}$-lipped $SU(5)$
GUT~\cite{Barr:1981qv, Derendinger:1983aj,Antoniadis:1987dx}, two pairs of TeV scale vector-like multiplets with origins
in $\mathcal{F}$-theory~\cite{Jiang:2006hf, Jiang:2009zza, Jiang:2009za,Li:2010dp} model building, and the boundary
conditions of No-Scale Supergravity~\cite{Cremmer:1983bf,Ellis:1983sf, Ellis:1983ei, Ellis:1984bm, Lahanas:1986uc}.
It appears that the No-Scale scenario, particularly vanishing of the Higgs
bilinear soft term $B_\mu$, comes into its own only when applied at an elevated
scale, approaching the Planck mass.  $M_{\cal F} \simeq 7\times 10^{17}$ GeV,
the point of the second stage $SU(5)\times U(1)_{\rm X}$ unification,
emerges in turn as a suitable candidate scale
only when substantially decoupled from the primary GUT scale unification
of $SU(3)_C\times SU(2)_L$ via the modification to the renormalization
group equations (RGEs) from the extra ${\cal F}$-theory vector multiplets.

In particular, we have systematically established the hyper-surface within 
the $\tan \beta$, top quark mass $m_{t}$, gaugino mass 
$M_{1/2}$, and vector-like particle mass $M_{V}$ parameter 
volume which is compatible with the application of the simplest
No-Scale SUGRA boundary conditions~\cite{Cremmer:1983bf,Ellis:1983sf, Ellis:1983ei, Ellis:1984bm, Lahanas:1986uc}, 
particularly the vanishing of the Higgs bilinear soft 
term $B_\mu$ at the ultimate ${\cal{F}}$-$SU(5)$ 
unification scale~\cite{Li:2010ws, Li:2010mi}. 
We have demonstrated that simultaneous adherence to all current experimental 
constraints, most importantly contributions to the muon anomalous 
magnetic moment $(g-2)_\mu$~\cite{Bennett:2004pv}, 
the branching ratio limit on 
$(b \rightarrow s\gamma)$~\cite{Barberio:2007cr, Misiak:2006zs}, 
and the 7-year WMAP relic density measurement~\cite{Spergel:2003cb, Spergel:2006hy, Komatsu:2010fb}, 
dramatically reduces the allowed solutions to a highly non-trivial 
``golden strip'' with $\tan \beta \simeq 15$, $m_{t} = 173.0-174.4 ~{\rm GeV}$, 
$M_{1/2} = 455-481 ~{\rm GeV}$, and $M_{V} = 691-1020 ~{\rm GeV}$, effectively 
eliminating all extraneously tunable model parameters,
where the consonance of the theoretically viable $m_{t}$ 
range with the experimentally established value~\cite{:2009ec} is an independently 
correlated ``postdiction''.
Finally, taking a fixed $Z$-boson mass, we have dynamically determined the universal
gaugino mass $M_{1/2}$ and fixed $\tan \beta$ via the ``Super No-Scale''
mechanism~\cite{Li:2010uu}, that being the secondary minimization, or {\it minimum minimorum},
of the minimum $V_{\rm min}$ of the Higgs potential for the electroweak symmetry breaking (EWSB) vacuum.

These models are moreover quite 
interesting from a phenomenological point of view~\cite{Jiang:2009zza,Jiang:2009za}. The predicted vector-like particles
can be observed at the Large Hadron Collider, and the partial lifetime for proton decay 
in the leading ${(e|\mu)}^{+} \pi^0 $ channels falls around 
$5 \times 10^{34}$ years, testable at the future 
Hyper-Kamiokande~\cite{Nakamura:2003hk} and
Deep Underground Science and Engineering 
Laboratory (DUSEL)~\cite{Raby:2008pd} experiments~\cite{Li:2009fq, Li:2010dp}.
The lightest CP-even Higgs boson mass can be increased~\cite{HLNT}, 
hybrid inflation can be naturally realized, and the 
correct cosmic primordial density fluctuations can be 
generated~\cite{Kyae:2005nv}.

%%%%%%%%%%%%%%%%%%%%%%%%%%%%%%%%%%%%%%%%%%%%%%%%%%%%%%%%%%%%%%%%%%%%%%%%%%%%%

\section{No-Scale Foundations of $\cal{F}$-$SU(5)$}

In the traditional framework, 
supersymmetry is broken in 
the hidden sector, and then its breaking effects are
mediated to the observable sector
via gravity or gauge interactions. In GUTs with
gravity mediated supersymmetry breaking, also known as the
minimal Supergravity (mSUGRA) model, 
the supersymmetry breaking soft terms can be parameterized
by four universal parameters: the gaugino mass $M_{1/2}$,
scalar mass $M_0$, trilinear soft term $A$, and
the ratio of Higgs VEVs $\tan \beta$ at low energy,
plus the sign of the Higgs bilinear mass term $\mu$.
The $\mu$ term and its bilinear 
soft term $B_{\mu}$ are determined
by the $Z$-boson mass $M_Z$ and $\tan \beta$ after
the electroweak (EW) symmetry breaking.

To solve the cosmological constant
problem, No-Scale Supergravity was proposed~\cite{Cremmer:1983bf,Ellis:1983sf, Ellis:1983ei, Ellis:1984bm, Lahanas:1986uc}. 
No-scale Supergravity is defined as the subset of Supergravity models
which satisfy the following three constraints~\cite{Cremmer:1983bf,Ellis:1983sf, Ellis:1983ei, Ellis:1984bm, Lahanas:1986uc}:
(i) the vacuum energy vanishes automatically due to the suitable
 K\"ahler potential; (ii) at the minimum of the scalar
potential, there are flat directions which leave the 
gravitino mass $M_{3/2}$ undetermined; (iii) the super-trace
quantity ${\rm Str} {\cal M}^2$ is zero at the minimum. Without this,
the large one-loop corrections would force $M_{3/2}$ to be either
zero or of Planck scale. A simple K\"ahler potential which
satisfies the first two conditions is
\begin{eqnarray} 
K &=& -3 \ln( T+\overline{T}-\sum_i \overline{\Phi}_i
\Phi_i)~,~
\label{NS-Kahler}
\end{eqnarray}
where $T$ is a modulus field and $\Phi_i$ are matter fields.
The third condition is model dependent and can always be satisfied in
principle~\cite{Ferrara:1994kg}.

The scalar fields of Eq.~(\ref{NS-Kahler}) 
parameterize the coset space
$SU(N_C+1, 1)/(SU(N_C+1)\times U(1))$, where $N_C$ is the number
of matter fields. Analogous structures appear in the 
$N\ge 5$ extended Supergravity theories~\cite{Cremmer:1979up}, for example,
$N_C=4$ for $N=5$, which can be realized in the compactifications
of string theory~\cite{Witten:1985xb, Li:1997sk}. 
The non-compact structure of the symmetry
implies that the potential is not only constant but actually
identical to zero.  For the simple example K\"ahler potential
given above, one can readily check that
the scalar potential is automatically positive semi-definite,
and has a flat direction along the $T$ field.  Likewise, it 
may be verified that the simplest No-Scale boundary conditions
$M_0=A=B_{\mu}=0$ emerge dynamically, while $M_{1/2}$ may be
non-zero at the unification scale, allowing for low energy SUSY breaking. 

The specific K\"ahler potential of Eq.~(\ref{NS-Kahler}) has been independently derived in both
weakly coupled heterotic string theory~\cite{Witten:1985xb} and the leading order compactification of
M-theory on $S^1/Z_2$~\cite{Li:1997sk}.  Note that in both cases, the Yang-Mills fields span a ten dimensional space-time.
It is not obtained directly out of F-theory,
as represented for example by the strong coupling lift from Type
IIB intersecting D-brane model building with D7- and
D3-branes~\cite{Beasley:2008dc,Beasley:2008kw, Donagi:2008ca, Donagi:2008kj},
where the Yang-Mills fields on the D7-branes occupy an eight dimensional space-time.
Nevertheless, it is certainly possible in principle to
calculate a gauge kinetic function, Kahler potential and superpotential
in the context of Type IIB interecting D-brane model building, and
the F-theory could thus admit a more general definition
of No-Scale Supergravity, as realized by a K\"ahler potential like
\begin{eqnarray}
K &=& -\ln(S + \overline{S}) -\ln(T_1 + \overline{T}_1)
\nonumber \\
&& -\ln(T_2 + \overline{T}_2) -\ln(T_3 + \overline{T}_3) \, ,
\label{Kahler2}
\end{eqnarray}
where only three of the moduli fields $S$ and $T_i$ may yield non-zero F-terms.

In Ref.~\cite{Giddings:2001yu}, No-Scale Supergravity was obtained in the
Type IIB and F-theory compactifications at the leading order.  Likewise,
the subsequently introduced KKLT~\cite{Kachru:2003aw} constructions also
manifest a No-Scale SUGRA structure at the classical level.
Indeed, the No-Scale features are generically obtained at
the tree-level in string theory compactifications due to
the presence of three complex extra dimensions.  However, this classical
level result is rather precariously balanced, and may be spoiled by
quantum corrections to the superpotential including flux contributions,
instanton effects, gaugino condensation, and the next order $\alpha'$
corrections.  In this sense, we consider the KKLT type SUGRA models as
a generalization or extension of the elemental No-Scale form.

The No-Scale ${\cal F}$-$SU(5)$ model under discussion has been constructed locally in
F-theory~\cite{Jiang:2009zza, Jiang:2009za}, although the mass of the additional vector-like multiplets,
and even the fact of their existence, is not mandated by the F-theory, wherein it is also possible to
realize models with only the traditional Flipped (or Standard) $SU(5)$ field content.  We claim only an inherent consistency
of their conceptual origin out of the F-theoretic construction, and take the manifest phenomenological benefits which accompany
the natural elevation of the secondary GUT unification phase to $M_{\cal{F}} \simeq 7 \times 10^{17}$~GeV as justification for the greater
esteem which we hold for this particular model above other alternatives.
There are, though, also delicate questions of compatibility between the local F-theoretic model building
origins and the purely field-theoretic RGE running which we employ up to the presumed high scale.
As one approaches the Planck mass $M_{\rm Pl}$, consideration must be given to the role which will be played by
Kaluza Klein (KK) and string mode excitations, and also to corrections of order $\alpha^\prime$ from stabilization of the
global volume of the six-dimensional internal space in association with the establishment of the
string scale $M_{\rm S} \propto (M_{\rm Pl} / R_{\rm global}^3)$.

The most important question is whether our model can in fact be embedded into a globally consistent framework.
Without such, we do not know the concrete K\"ahler potential of the SM fermions and Higgs fields,
and cannot by this means explicitly calculate the supersymmetry breaking scalar masses and
trilinear soft terms.  This construction remains elusive though, and is beyond the reach of the current
work.  Regardless, one may anticipate that in such a globally consistent model, a string
scale of order $10^{17}$~GeV would indeed be realized, as in the weakly coupled
heterotic string theory, tying in nicely with our na\"{\i}vely projected value for $M_{\cal F}$.
It seems additionally that a field-theoretic application of the No-Scale boundary conditions might
prove to be validated in this case.  Moreover, we would not necessarily require the presence of
instanton effects or gaugino condensation for stabilization of the modulus $T$ as in the KKLT mechanism.
This is crucial, because such effects can have the negative side effect of destroying
the leading No-Scale structure.  In fact, we could have no gaugino condensation
at all, or the superpotential from gaugino condensation might only depend on $S$, as again exemplified
in the Type IIB intersecting D-brane models~\cite{Blumenhagen:2005mu}.

Such considerations, coupled with the demonstrated testability and phenomenological success of the first order
analysis in the simplest No-Scale SUGRA framework, argue for a continuing study of the generalized
No-Scale SUGRA picture.  It is important to note that there exist several such generalizations,
including the previously mentioned Type II intersecting D-brane models~\cite{Cvetic:2004ui, Chen:2007px, Chen:2007zu},
mirage mediation of flux compactifications~\cite{Choi:2004sx, Choi:2005ge},
and the extraction of SUSY breaking soft terms from the leading order compactification of M-theory
on $S^1/Z_2$~\cite{Nilles:1997cm, Nilles:1998sx, Lukas:1997fg, Lukas:1998yy, Li:1998rn};
in the latter case we have previously obtained (in a different model context) a generalization employing
modulus dominated SUSY breaking~\cite{Li:1998rn}.

In this paper, however, we maintain a ``first steps first'' perspective, concentrating on the simplest No-Scale
Supergravity and reserving any such extensions for the future.
The potential for stringy modifications duly noted, we then essentially aim to study an F-theory
{\it inspired} variety of low energy SUSY phenomenology, remaining agnostic
as to the details of the K\"ahler structure.  Nevertheless, by studying the simplest No-Scale
Supergravity, we may still expect to encapsulate the correct leading order behavior.
We likewise maintain the simplicity of a leading order approximation by neglecting consideration of
any stringy threshold corrections, the substantive onset of which is anyway expected to be deferred to
$M_{\cal F}$, the true GUT scale of this model. It should be added that since the running of the gauge
couplings is logarithmically dependent upon the mass scale, the contributions
to the RGEs from the string and KK mode excitations are quite small.

%%%%%%%%%%%%%%%%%%%%%%%%%%%%%%%%%%%%%%%%%%%%%%%%%%%%%%%%%%%%%%%%%%%%%%%%%%%%%

\section{ The Super No-Scale Mechanism }

The single relevant modulus field in the simplest 
stringy No-Scale Supergravity is the K\"ahler
modulus $T$, a characteristic of the Calabi-Yau manifold,
the dilaton coupling being irrelevant.
We consider the gaugino mass $M_{1/2}$ as a useful modulus
related to the F-term of $T$, stipulating, in other words,
that the gauge kinetic function must depend on $T$.  This is realized,
for example, in the Type IIB intersecting D-brane models~\cite{Blumenhagen:2005mu}
where gauge kinetic functions explicitly depend on both $S$ and $T_i$, as in Eq.~(\ref{Kahler2}).
Again, since the F-theory may be considered as a strongly coupled
formulation of the Type IIB string theory, it is natural to believe
that the gauge kinetic function under this lift depends on $T$ as well.
While the limit is quite suggestive, lacking still a concrete
globally consistent embedding, we cannot definitively prove that
the superpotential remains unperturbed by $T$.

Proceeding tentatively as such, the F-term of $T$ generates the gravitino mass $M_{3/2}$,
which is proportionally equivalent to $M_{1/2}$.
Exploiting the simplest No-Scale boundary condition at $M_{\cal F}$ and 
running from high energy to low energy under the RGEs,
there can be a secondary minimization, or {\it minimum minimorum}, of the minimum of the
Higgs potential $V_{\rm min}$ for the EWSB vacuum.
Since $V_{\rm min}$ depends on $M_{1/2}$, the gaugino mass $M_{1/2}$ is consequently 
dynamically determined by the equation $dV_{\rm min}/dM_{1/2}=0$,
aptly referred to as the ``Super No-Scale'' mechanism~\cite{Li:2010uu}.

It could easily have been that in consideration of the above technique, there were: A) too few undetermined parameters,
with the $B_{\mu}=0$ condition forming an incompatible over-constraint, and thus demonstrably false, or B) so many
undetermined parameters that the dynamic determination possessed many distinct solutions, or was so far separated
from experiment that it could not possibly be demonstrated to be true.  The actual state of affairs is much
more propitious, being specifically as follows. The three parameters $M_0,A,B_{\mu}$ are once again identically zero at the
boundary because of the defining K\"ahler potential, and are thus known at all other scales as well by the RGEs.  The
minimization of the Higgs scalar potential with respect to the neutral elements of both SUSY Higgs doublets gives two
conditions, the first of which fixes the magnitude of $\mu$.  The second condition, which would traditionally be used
to fix $B_{\mu}$, instead here enforces a consistency relationship on the remaining parameters, being that
$B_{\mu}$ is already constrained.

In general, the $B_{\mu} = 0$ condition gives a hypersurface of solutions cut out from a very large parameter space.
If we lock all but one parameter, it will give the final value.  If we take a slice of two dimensional space, as has been 
described, it will give a relation between two parameters for all others fixed.
In a three-dimensional view with $B_{\mu}$ on the vertical axis, this
curve is the ``flat direction'' line along the bottom of the trench of $B_{\mu}=0$ solutions.  In general, we
must vary at least two parameters rather than just one in isolation, in order that their mutual compensation may transport
the solution along this curve.  The most natural first choice is in some sense the pair of prominent unknown inputs 
$M_{1/2}$ and $\tan \beta$, as was demonstrated in Ref.~\cite{Li:2010uu}.

Having come to this point, it is by no means guaranteed that the potential
will form a stable minimum.  It must be emphasized that the $B_{\mu}=0$ No-Scale
boundary condition is the central agent affording this determination, as it is the extraction of the parameterized
parabolic curve of solutions in the two compensating variables which allows for a localized, bound nadir point to be
isolated by the Super No-Scale condition, dynamically determining {\it both} parameters.  The background surface of
$V_{\rm min}$ for the full parameter space outside the viable $B_{\mu}=0$ subset is, in contrast, a steadily inclined
and uninteresting function.  In our prior study, the local {\it minimum minimorum} of $V_{\rm min}$ for the
choices $M_{V}=1000$ GeV and $m_{t}=173.1$ GeV dynamically established $M_{1/2} \simeq 450~{\rm GeV}$, and
$\tan \beta \simeq 15-20$.
%We reiterate and nominally update that result in Fig.~\cite{fig:dV_MZ_tanb}.
Although we have remarked that $M_{1/2}$
and $\tan \beta$ have no {\it directly} established experimental values, they are severely indirectly constrained by
phenomenology in the context of this model~\cite{Li:2010ws,Li:2010mi}.  It is highly non-trivial that there should be
accord between the top-down and bottom-up perspectives, but this is indeed precisely what has been observed~\cite{Li:2010uu}. 

%%%%%%%%%%%%%%%%%%%%%%%%%%%%%%%%%%%%%%%%%%%%%%%%%%%%%%%%%%%%%%%%%%%%%%%%%%%%%

\section{ The GUT Higgs Modulus }

An alternate pair of parameters for which one may attempt to isolate a $B_{\mu} = 0$ curve,
which we consider for the first time in this work, is that of $M_{1/2}$ and the GUT scale $M_{32}$,
at which the $SU(3)_C$ and $SU(2)_L$ couplings initially meet.  Fundamentally, the latter corresponds
to the modulus which sets the total magnitude of the GUT Higgs field's VEVs.
$M_{32}$ could of course in some sense be considered a ``known'' quantity, taking the low energy couplings as input.
Indeed, starting from the measured SM gauge couplings and fermion Yukawa couplings at the standard 
$91.187$~GeV electroweak scale, we may calculate both $M_{32}$ and the final unification scale $M_{\cal F}$,
and subsequently the unified gauge coupling and SM fermion Yukawa couplings at $M_{\cal F}$, via
running of the RGEs.  However, since the VEVs of the GUT Higgs fields $H$ and $\overline{H}$ are considered
here as free parameters, the GUT scale $M_{32}$ must not be fixed either. As a consequence,
the low energy SM gauge couplings, and in particular the $SU(2)_L$ gauge coupling $g_2$,
will also run freely via this feedback from $M_{32}$.

We consider this conceptual release of a known quantity, in order to establish the
nature of the model's dependence upon it, to be a valid and valuable technique, and have employed it previously
with specific regards to ``postdiction'' of the top quark mass value~\cite{Li:2010mi}.  Indeed, forcing the
theoretical {\it output} of such a parameter is only possible in a model with highly constrained physics, and it
may be expected to meet success only by intervention of either grand coincidence or grand conspiracy of Nature.
Simultaneous to the recognition of the presence of a second dynamic modulus, we lock down the value of $\mu$,
which by contrast is a simple numerical parameter, and
ought then to be treated in a manner consistent with the top quark and vector-like mass parameters.
For this study, we choose a vector-like particle mass $M_V=1000$~GeV, and use the experimental top
quark mass input $m_{t}=173.1$~GeV.  We emphasize that the choice of $M_V=1000~{\rm GeV}$ is not an arbitrary
one, since a prior analysis~\cite{Li:2010mi} has shown that a $1$~TeV vector-like mass is in compliance
with all current experimental data and the No-Scale $B_{\mu}$=0 requirement.  The constant parameter $\mu$ is
set consistent with its value prior to the variation of the GUT modulus. 

\begin{figure}[ht]
        \centering
        \includegraphics[width=0.50\textwidth]{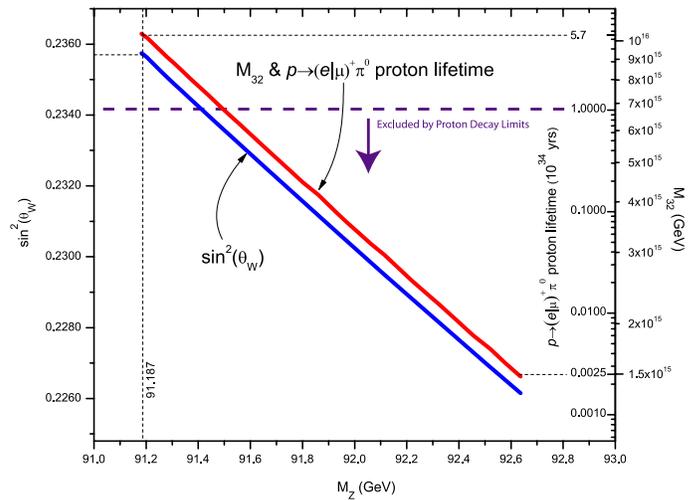}
        \caption{The interrelated variation of $\sin^2 (\theta_{\rm W})$, the GUT scale $M_{32}$ (logarithmic axis),
and the $Z$-boson mass $M_Z$ is demonstrated for the parameter strips which preserve $B_{\mu}=0$ and $\mu=460$~GeV at $M_{\cal F}$.
The variation in $M_Z$ is linked dominantly to motion of the EW couplings via $\sin^2 (\theta_{\rm W})$.
Also shown is the corresponding predicted proton lifetime in the leading
${(e|\mu)}^{+} \pi^0 $ channels, in units of $10^{34}$ years, with the current lower bound of $1.0 \times 10^{34}$ years indicated
by the dashed horizontal purple line.}
\label{fig:sin2T_MZ_M32}
\end{figure}

In actual practice, the variation of $M_{32}$ is achieved in the reverse by programmatic variation of the
Weinberg angle, holding the strong and electromagnetic couplings at their physically measured values.
Figure~\ref{fig:sin2T_MZ_M32} demonstrates the scaling between $\sin^2 (\theta_{\rm W})$,
$M_{32}$ (logarithmic axis), and the $Z$-boson mass.  The variation of $M_Z$ is attributed primarily to the motion
of the electroweak couplings, the magnitude of the Higgs VEV being held essentially constant.
We ensure also that the unified gauge coupling, SM fermion Yukawa couplings, and specifically also
the Higgs bilinear term $\mu \simeq 460$~GeV, are each held stable at the scale $M_{\cal F}$ to correctly
mimic the previously described procedure.

\begin{figure}[ht]
	\centering
	\includegraphics[width=0.50\textwidth]{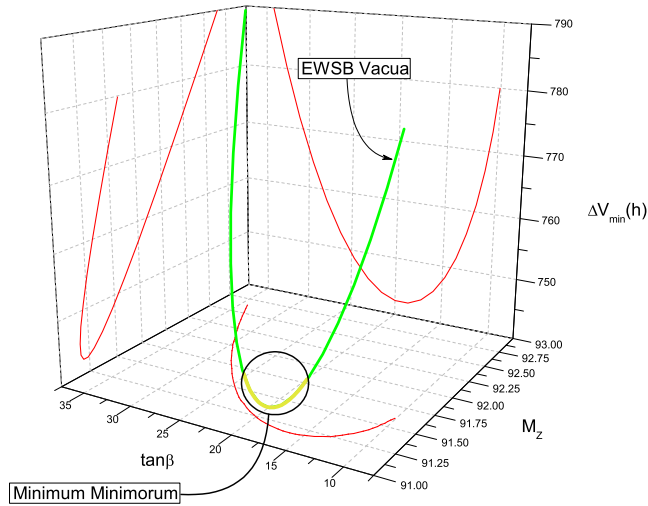}
	\caption{Three-dimensional graph of $(M_{Z},\tan \beta,\Delta V_{min}(h)$) space (green curve). The projections onto the three mutually perpendicular planes (red curves) are likewise shown. $M_{Z}$ and $\Delta V_{min}(h)$ are in units of GeV.  The dynamically preferred region, allowing for plausible variation, is circled and tipped in gold.}
\label{fig:dV_MZ_tanb}
\end{figure}

\begin{figure}[ht]
	\centering
	\includegraphics[width=0.50\textwidth]{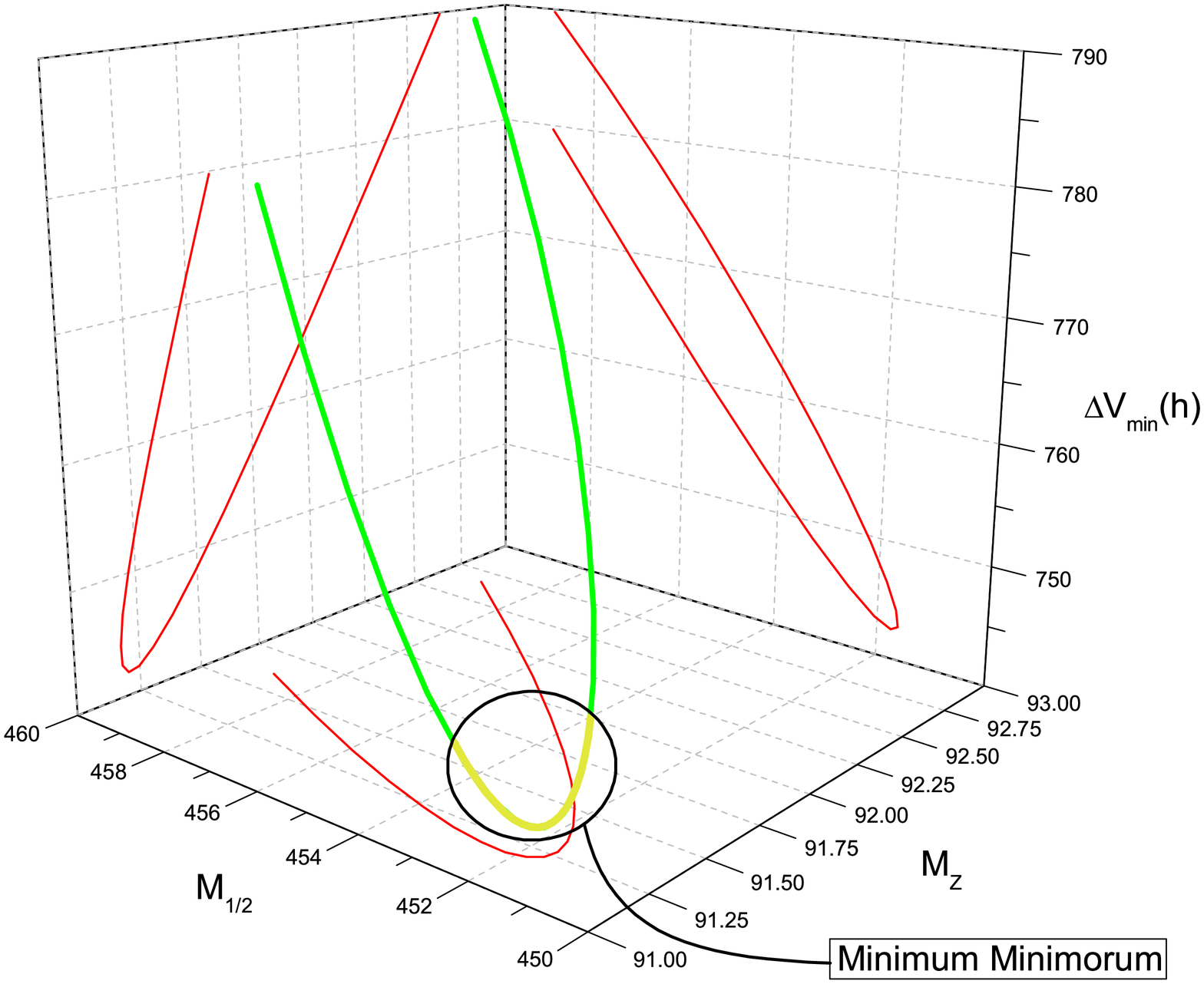}
	\caption{Three-dimensional graph of $(M_{Z},M_{1/2},\Delta V_{min}(h))$ space (green curve). The projections onto the three mutually perpendicular planes (red curves) are likewise shown. $M_{Z}$, $M_{1/2}$, and $\Delta V_{min}(h)$ are in units of GeV.  The dynamically preferred region, allowing for plausible variation, is circled and tipped in gold.}
\label{fig:dV_M12_MZ}
\end{figure}

\begin{figure}[ht]
	\centering
	\includegraphics[width=0.50\textwidth]{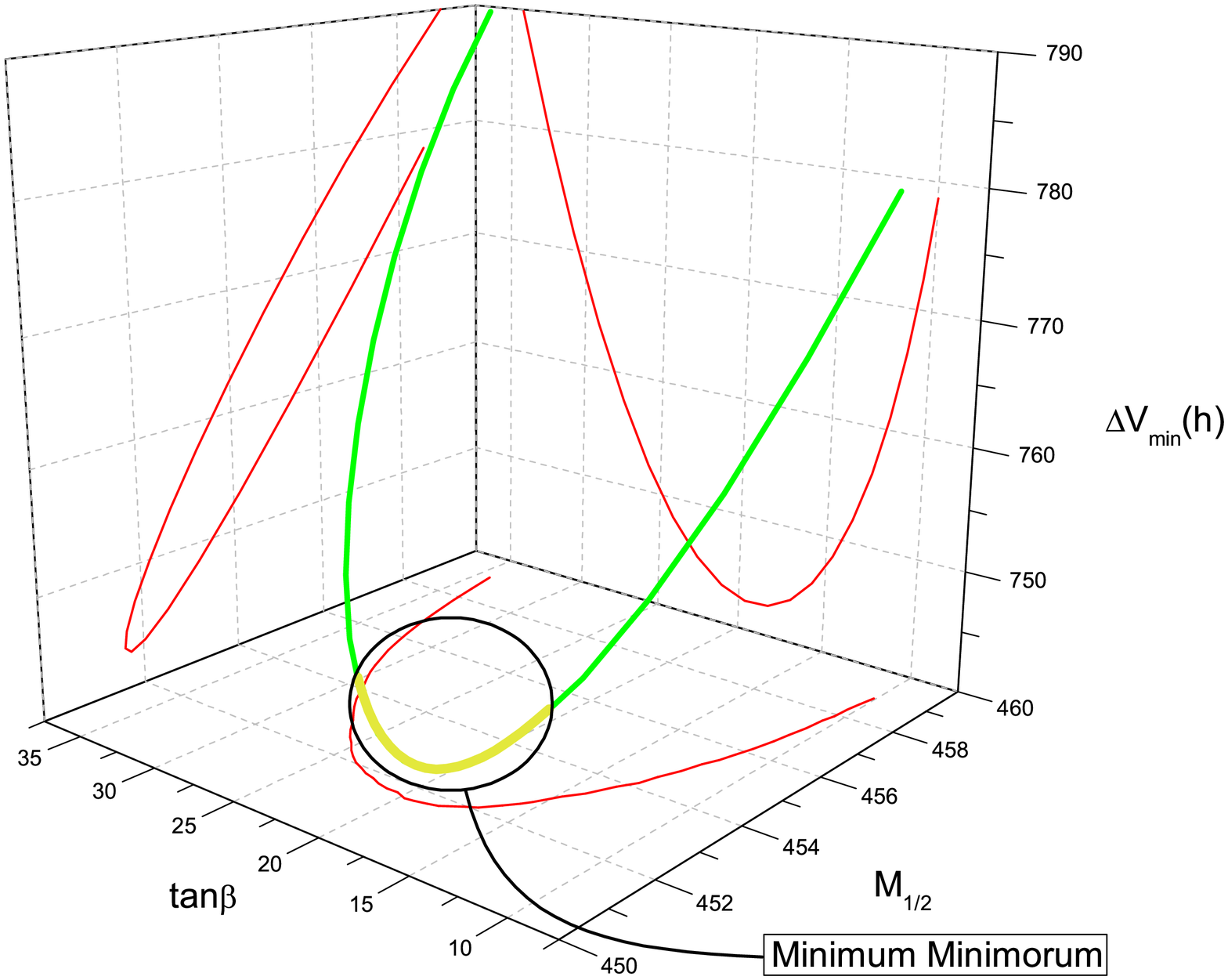}
	\caption{Three-dimensional graph of $(M_{1/2},\tan \beta,\Delta V_{min}(h)$) space (green curve). The projections onto the three mutually perpendicular planes (red curves) are likewise shown. $M_{1/2}$ and $\Delta V_{min}(h)$ are in units of GeV.  The dynamically preferred region, allowing for plausible variation, is circled and tipped in gold.}
\label{fig:dV_M12_tanb}
\end{figure}

\begin{figure}[ht]
	\centering
	\includegraphics[width=0.50\textwidth]{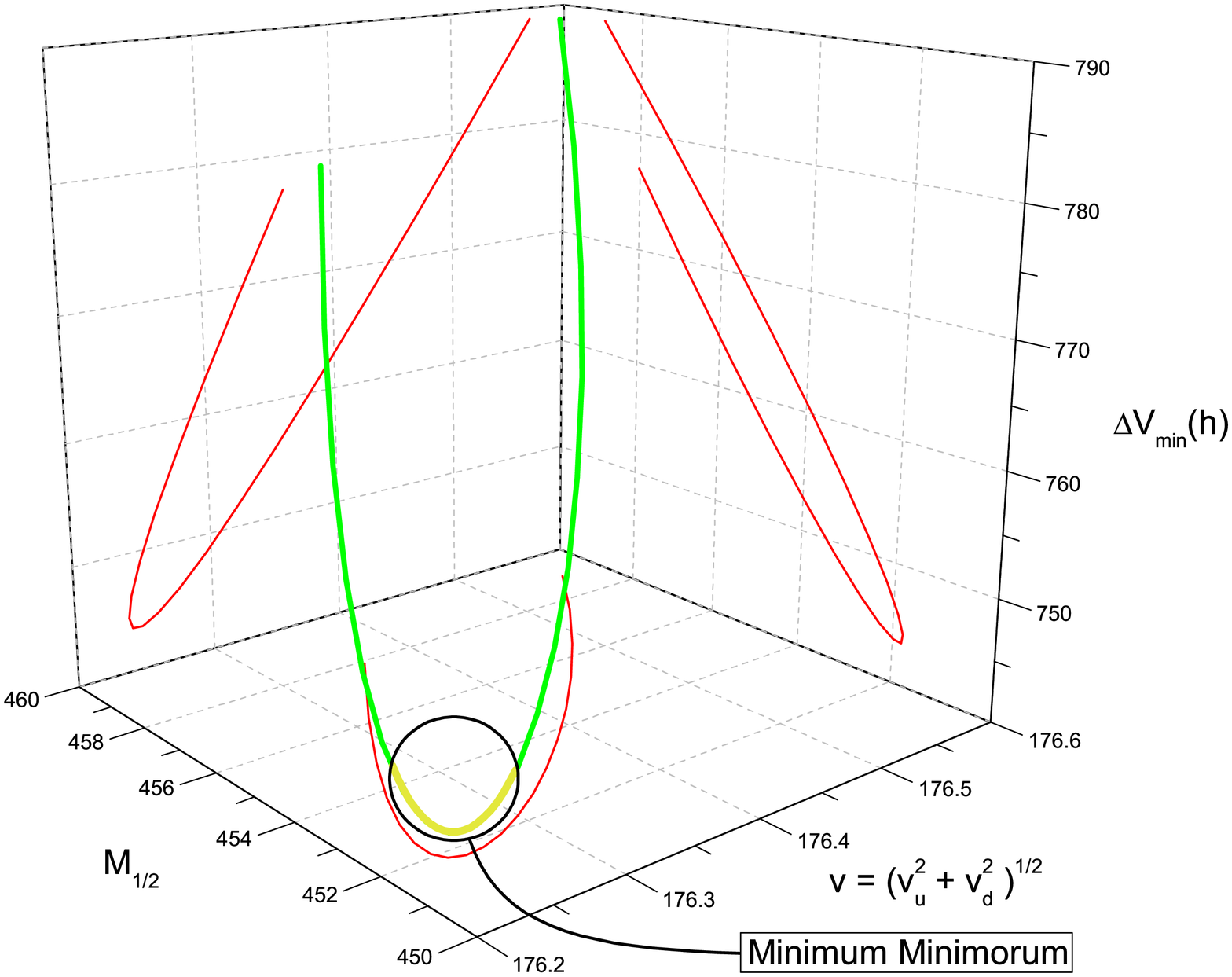}
	\caption{Three-dimensional graph of $(v,M_{1/2},\Delta V_{min}(h))$ space (green curve). The projections onto the three mutually perpendicular planes (red curves) are likewise shown. $M_{1/2}$, $v$, and $\Delta V_{min}(h)$ are in units of GeV.  The dynamically preferred region, allowing for plausible variation, is circled and tipped in gold.}
\label{fig:dV_M12_v}
\end{figure}

\begin{figure}[ht]
	\centering
	\includegraphics[width=0.50\textwidth]{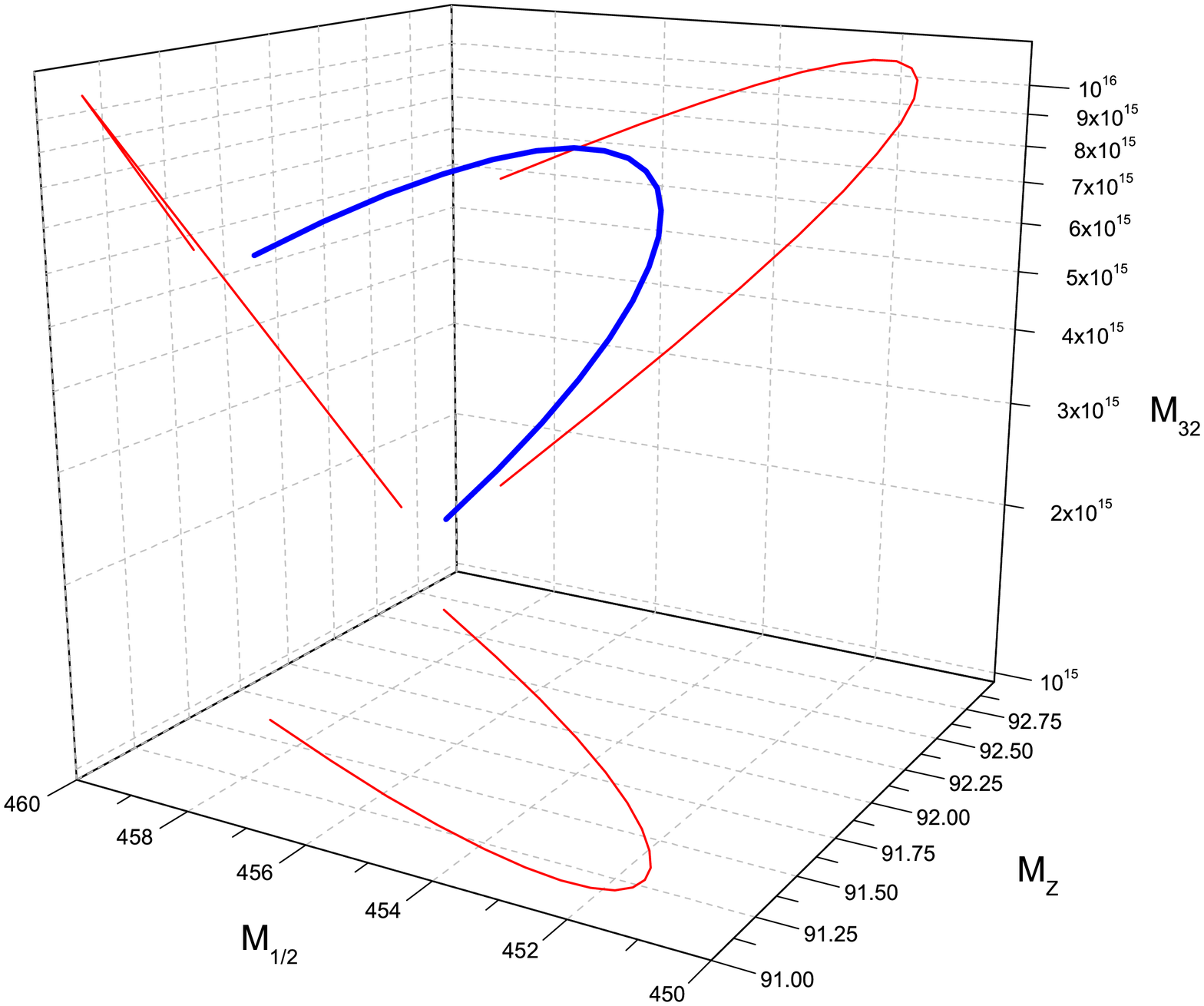}
	\caption{Three-dimensional graph of $(M_{Z},M_{1/2},M_{32})$ space (blue curve). The projections onto the three mutually perpendicular planes (red curves) are likewise shown. $M_{Z}$, $M_{1/2}$, and $M_{32}$ are in units of GeV.}
\label{fig:M32_M12_MZ}
\end{figure}

\begin{figure}[ht]
	\centering
	\includegraphics[width=0.50\textwidth]{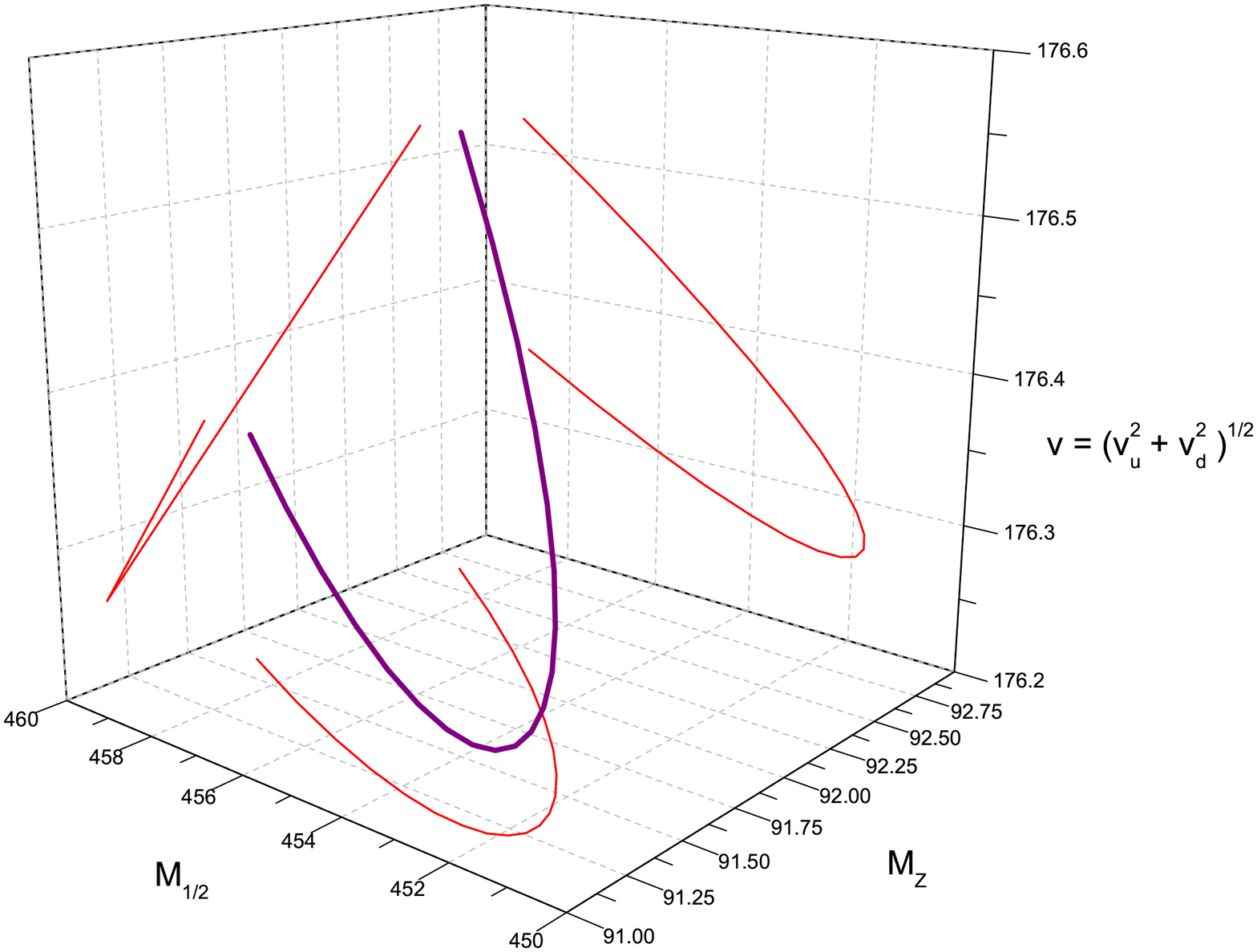}
	\caption{Three-dimensional graph of $(M_{Z},M_{1/2},v)$ space (purple curve). The projections onto the three mutually perpendicular planes (red curves) are likewise shown. $M_{Z}$, $M_{1/2}$, and $v$ are in units of GeV.}
\label{fig:v_M12_MZ}
\end{figure}

\begin{figure}[ht]
	\centering
	\includegraphics[width=0.50\textwidth]{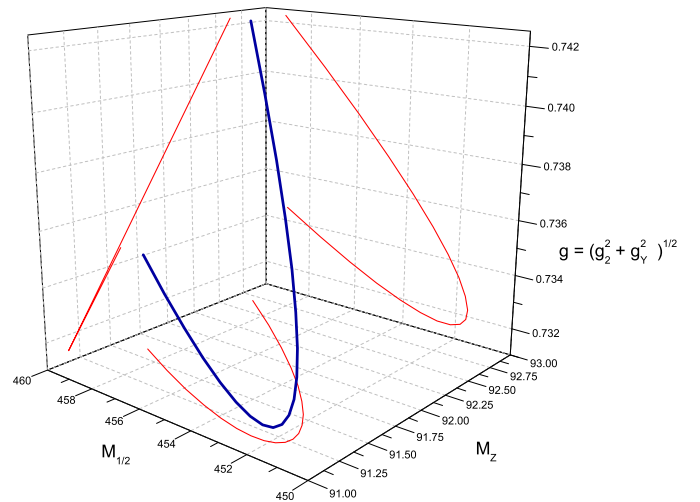}
	\caption{Three-dimensional graph of $(M_{Z},M_{1/2},g)$ space (royal blue curve). The projections onto the three mutually perpendicular planes (red curves) are likewise shown. $M_{Z}$ and $M_{1/2}$ are in units of GeV.}
\label{fig:g_M12_MZ}
\end{figure}

The parameter ranges for the variation depicted in Fig.~\ref{fig:sin2T_MZ_M32} are $M_Z = 91.18 - 92.64$,
$\sin^2(\theta_{\rm W}) = 0.2262 - 0.2357$, and $M_{32} = 1.5\times 10^{15} - 1.04 \times 10^{16}$~GeV, and likewise
also the same for Figs.~(\ref{fig:dV_MZ_tanb}-\ref{fig:g_M12_MZ}),
which will feature subsequently.  The {\it minimum minimorum} falls at the boundary of the prior list, dynamically
fixing $M_{32} \simeq 1.0 \times 10^{16}$~GeV and placing $M_{1/2}$ again in the vicinity of $450$~GeV.
The low energy SM gauge couplings are simultaneously constrained by means of the associated Weinberg angle, with
$\sin^2 (\theta_{\rm W}) \simeq 0.236$, in excellent agreement with experiment.
The corresponding range of predicted proton lifetimes in the leading ${(e|\mu)}^{+} \pi^0 $ modes is
$2.5\times 10^{31} - 5.7\times10^{34}$~years~\cite{Li:2010dp}.  If the GUT scale $M_{32}$ becomes excessively light, below about
$7 \times 10^{15}$~GeV, then proton decay would be more rapid than allowed by 
the recently updated lower bound of $1.0 \times 10^{34}$~years from Super-Kamiokande~\cite{:2009gd}.

We are cautious against making a claim in precisely the same vein for the dynamic determination of $M_Z \simeq 91.2$~GeV, since again
the crucial electroweak Higgs VEV is not a substantial element of the variation.  However, in {\it conjunction} with the radiative electroweak
symmetry breaking~\cite{Ellis:1982wr, Ellis:1983bp} numerically implemented within the {\tt SuSpect 2.34} code base~\cite{Djouadi:2002ze},
the fixing of the Higgs VEV and the determination of the electroweak scale may also plausibly be considered 
legitimate dynamic output, {\it if} one posits the $M_{F}$ scale input to be available {\it a priori}.

 By extracting a constant $\mu$ slice of the $V_{\rm min}$ hyper-surface, the secondary minimization condition on
$\tan \beta$ is effectively rotated, albeit quite moderately, relative to the procedure of Ref.~(\cite{Li:2010uu}).
The present minimization, referencing $M_{1/2}$, $M_{32}$ and $\tan \beta$, is again dependent upon $M_{V}$ and
$m_{t}$, while the previously described~\cite{Li:2010uu} determination of $\tan \beta$ was, by contrast, $M_{V}$ and $m_{t}$ invariant.
Recognizing that a minimization with all three parameters simultaneously active is required to declare all three
parameters to have been simultaneously dynamically determined, we emphasize the mutual consistency of the results.
We again stress that the new {\it minimum minimorum} is also consistent with the previously advertised
golden strip, satisfying all presently known experimental constraints to our available resolution.
It moreover also addresses the problems of the SUSY breaking scale and gauge hierarchy~\cite{Li:2010uu},
insomuch as $M_{1/2}$ is determined dynamically.

%%%%%%%%%%%%%%%%%%%%%%%%%%%%%%%%%%%%%%%%%%%%%%%%%%%%%%%%%%%%%%%%%%%%%%%%%%%%%

\section{ The Minimum Minimorum of the Electroweak Higgs Potential }

In supersymmetric SMs, there is a pair of Higgs doublets
$H_u$ and $H_d$ which give mass to the up-type quarks and
down-type quarks/charged leptons, respectively. The one-loop effective
Higgs potential in the 't Hooft-Landau gauge and in the $\overline{\rm DR}$
scheme is given by
\begin{eqnarray}
V_{\rm eff} &=& V_0(H_u^0,~H_d^0) + V_1(H_u^0,~H_d^0)~,~\,
\end{eqnarray}
where 
\begin{eqnarray}
V_0&=& (\mu^2 + m_{H_u}^2) (H_u^0)^2 + (\mu^2 + m_{H_d}^2) (H_d^0)^2
\nonumber \\ &&
-2 B_{\mu} \mu H_u^0 H_d^0 + {\frac{g_2^2 + g_Y^2}{8}} \left[(H_u^0)^2-(H_d^0)^2\right]^2
 ~,~\,
\end{eqnarray}
\begin{eqnarray}
V_1 &=&  \sum_i {\frac{n_i}{64\pi^2}} m_i^4(\phi)
\left( {\rm ln}{{\frac{m_i^2(\phi)}{Q^2}} -{\frac{3}{2}}} \right) ~,~\,
\end{eqnarray}
where $m_{H_u}^2$ and $m_{H_d}^2$ are the supersymmetry breaking soft masses,
$g_2$ and $g_Y$ are respectively the gauge couplings of $SU(2)_L$ and
$U(1)_Y$, $n_i$ and $m_i^2(\phi)$ are respectively the degree of freedom and mass
for $\phi_i$, and $Q$ is the renormalization scale.
In our numerical results in the figures, we shall designate differences in
the fourth-root of the effective Higgs potential as $\Delta V_{min}(h)\equiv V_{\rm eff}^{1/4}$,
measured in units of GeV relative to an arbitrary overall zero-offset.

We have revised the {\tt SuSpect 2.34} 
code base~\cite{Djouadi:2002ze} to incorporate our specialized
No-Scale ${\cal F}$-$SU(5)$ with vector-like mass algorithm, and accordingly employ
two-loop RGE running for the SM gauge couplings, and
one-loop RGE running for the SM fermion Yukawa couplings,
$\mu$ term, and SUSY breaking soft terms. For our choice of
$M_V=1000$ GeV, $m_t=173.1$ GeV, and $\mu(M_{\cal F}) \simeq 460$ GeV, 
we present the one-loop effective
Higgs potential $\Delta V_{min}(h)$ in terms of $M_{Z}$ and $\tan \beta$ in
Fig.~\ref{fig:dV_MZ_tanb}, in terms of 
$M_{Z}$ and $M_{1/2}$ in
Fig.~\ref{fig:dV_M12_MZ},
in terms of 
$M_{1/2}$ and tan$\beta$ in
Fig.~\ref{fig:dV_M12_tanb},
and in terms of $v$ and $M_{1/2}$ in
Fig.~\ref{fig:dV_M12_v}, where $v=\sqrt{v_{u}^{2}+v_{d}^{2}}$, 
$v_u=\langle H^0_u \rangle$, and $v_d=\langle H^0_d \rangle$.
These figures clearly demonstrate the localization
of the {\it minimum minimorum} of the Higgs potential,
corroborating the dynamical determination of $\tan \beta \simeq 15-20$
and $M_{1/2} \simeq 450$~GeV in~\cite{Li:2010uu}.

Additionally, we exhibit the $(M_{Z},M_{1/2},M_{32})$ space in Fig.~\ref{fig:M32_M12_MZ},
the $(M_{Z},M_{1/2},v)$ space in Fig.~\ref{fig:v_M12_MZ},
and the $(M_{Z},M_{1/2},g)$ space in Fig.~\ref{fig:g_M12_MZ},
where $g=\sqrt{g_{2}^{2}+g_{Y}^{2}}$. Fig.~\ref{fig:M32_M12_MZ} demonstrates that $M_{32} \simeq 1.0 \times 10^{16}$~GeV
at the {\it minimum minimorum}, which correlates to $M_{Z} \simeq 91.2$~GeV,
or more directly, $\sin^2 (\theta_{\rm W}) \simeq 0.236$.
Together, the alternate perspectives of Figs.~\ref{fig:M32_M12_MZ},
~\ref{fig:v_M12_MZ},
and~\ref{fig:g_M12_MZ} complete the view given in Figs.~\ref{fig:dV_MZ_tanb},~\ref{fig:dV_M12_MZ},~\ref{fig:dV_M12_tanb},
and~\ref{fig:dV_M12_v}
to visually tell the story of the dynamic interrelation between the $M_Z$, $M_{1/2}$, and $M_{32}$ scales,
as well as the electroweak gauge couplings, and the Higgs VEVs.
The curves in each of these figures represent only those points that satisfy the $B_{\mu}$ = 0 requirement, as dictated by No-Scale
Supergravity, serving as a crucial constraint on the dynamically determined parameter space. Ultimately, it is the significance of
the $B_{\mu} = 0$ requirement that separates the No-Scale ${\cal F}$-$SU(5)$ with vector-like particles from the entire
compilation of prospective string theory derived models. By means of the $B_{\mu}$ = 0 vehicle, No-Scale ${\cal 
 F}$-$SU(5)$ has surmounted the paramount challenge of phenomenology, that of dynamically determining the electroweak scale,
the scale of fundamental prominence in particle physics.

We wish to note that recent progress has been made in incorporating more precise numerical calculations into
our baseline algorithm for No-Scale ${\cal F}$-$SU(5)$ with vector-like particles. Initially, when we commenced the task of fully
developing the phenomenology of this model, the extreme complexity of properly numerically implementing No-Scale
${\cal F}$-$SU(5)$ with vector-like particles compelled a gradual strategy for construction and persistent enhancement of the
algorithm. Preliminary findings of a precision improved algorithm indicate that compliance with the 7-year WMAP relic density
constraints requires a slight upward shift to $\tan \beta \simeq 19-20$ from the value computed in Ref.~\cite{Li:2010ws},
suggesting a potential convergence to even finer resolution of the dynamical determination of $\tan \beta$ given by the Super No-Scale
mechanism, and the value demanded by the experimental relic density measurements.  We shall furnish a comprehensive analysis of the
precision improved algorithm at a later date.

%%%%%%%%%%%%%%%%%%%%%%%%%%%%%%%%%%%%%%%%%%%%%%%%%%%%%%%%%%%%%%%%%%%%%%%%%%%%%

\section{ Probing The Blueprints of the No-Scale Multiverse at the Colliders }

We offer in closing a brief summary of direct collider, detector, and telescope level tests
which may probe the blueprints of the No-Scale Multiverse which we have laid out.  As to the deep
question of whether the ensemble be literal in manifestation, or merely the conceptual
superset of unrealized possibilities of a single island Universe, we pretend no definitive
answer.  However, we have argued that the emergence {\it ex nihilo} of seedling universes which fuel
an eternal chaotic inflation scenario is particularly plausible, and even natural, within No-Scale
Supergravity, and our goal of probing the specific features of our own Universe which might
implicate its origins in this construction are immediately realizable and practicable.

The unified gaugino $M_{1/2}$ at the unification scale $M_{\cal F}$ can be reconstructed
from impending LHC events by determining the gauginos $M_{1}$, $M_{2}$, and $M_{3}$ at the
electroweak scale, which will in turn require knowledge of the masses for the neutralinos, charginos,
and the gluino. Likewise, $\tan \beta$ can be ascertained in principle from a distinctive experimental
observable, as was accomplished for mSUGRA in~\cite{Arnowitt:2008bz}.  We will not undertake a comprehensive
analysis here of the reconstruction of $M_{1/2}$ and $\tan \beta$, but will offer for now a cursory examination
of typical events expected at the LHC. We leave the detailed compilation of the experimental
observables necessary for validation of the No-Scale ${\cal F}$-$SU(5)$ at the LHC for the future,
and we especially encourage those specializing in such research to investigate the No-Scale
${\cal F}$-$SU(5)$.

For the benchmark SUSY spectrum presented in Table~\ref{tab:masses}, we have adopted
the specific values $M_{1/2}=453$, $\tan \beta=15$ and $M_Z=91.187$.
We expect that higher order corrections will shift the precise location of the {\it minimum minimorum}
a little bit, for example, within the encircled gold-tipped regions of the diagrams in the prior section.
We have selected a ratio for $\tan \beta$ at the lower end of this range for consistency with our previous
study~\cite{Li:2010uu}, and to avoid stau dark matter.

\begin{table}[ht]
  \small
	\centering
	\caption{Spectrum (in GeV) for the benchmark point. 
Here, $M_{1/2}$ = 453 GeV, $M_{V}$ = 1000 GeV, $m_{t}$ = 173.1 GeV, $M_{Z}$ = 91.187 GeV, $\mu (M_{\cal F})$ = 460.3 GeV, $\Delta V_{min}(h)$ = 748 GeV, $\Omega_{\chi}$ = 0.113, $\sigma_{SI} = 2 \times 10^{-10}$ pb, and $\left\langle \sigma v \right\rangle_{\gamma\gamma} = 1.8 \times 10^{-28} ~cm^{3}/s$. The central prediction for the $p \!\rightarrow\! {(e\vert\mu)}^{\!+}\! \pi^0$ proton lifetime is around $4.9 \times 10^{34}$ years. The lightest neutralino is 99.8\% Bino.}

		\begin{tabular}{|c|c||c|c||c|c||c|c||c|c||c|c|} \hline		
    $\widetilde{\chi}_{1}^{0}$&$94$&$\widetilde{\chi}_{1}^{\pm}$&$184$&$\widetilde{e}_{R}$&$150$&$\widetilde{t}_{1}$&$486$&$\widetilde{u}_{R}$&$947$&$m_{h}$&$120.1$\\ \hline
    $\widetilde{\chi}_{2}^{0}$&$184$&$\widetilde{\chi}_{2}^{\pm}$&$822$&$\widetilde{e}_{L}$&$504$&$\widetilde{t}_{2}$&$906$&$\widetilde{u}_{L}$&$1032$&$m_{A,H}$&$916$\\ \hline
     $\widetilde{\chi}_{3}^{0}$&$817$&$\widetilde{\nu}_{e/\mu}$&$498$&$\widetilde{\tau}_{1}$&$104$&$\widetilde{b}_{1}$&$855$&$\widetilde{d}_{R}$&$988$&$m_{H^{\pm}}$&$921$\\ \hline
    $\widetilde{\chi}_{4}^{0}$&$821$&$\widetilde{\nu}_{\tau}$&$491$&$\widetilde{\tau}_{2}$&$499$&$\widetilde{b}_{2}$&$963$&$\widetilde{d}_{L}$&$1035$&$\widetilde{g}$&$617$\\ \hline
		\end{tabular}
		\label{tab:masses}
\end{table}

At the benchmark point, we calculate
$\Omega_{\chi} = 0.113$ for the cold dark matter relic density. 
The phenomenology is moreover consistent with the LEP limit on the lightest CP-even Higgs boson
mass, $m_{h} \geq 114$ GeV~\cite{Barate:2003sz,Barate:2003sz}, the CDMSII~\cite{Ahmed:2008eu} and
Xenon100~\cite{Aprile:2010um} upper limits on the spin-independent cross section
$\sigma_{SI}$, and the Fermi-LAT space telescope constraints~\cite{Abdo:2010dk} on the 
photon-photon annihilation cross section $\left\langle \sigma v \right\rangle_{\gamma\gamma}$.
The differential cross-sections and branching ratios have been calculated with
{\tt PGS4}~\cite{PGS4} executing a call to {\tt PYTHIA 6.411}~\cite{Sjostrand:2006za}, using our
specialized No-Scale algorithm integrated into the {\tt SuSpect 2.34} code for initial computation of
the sparticle masses.

The benchmark point resides in the region of the experimentally allowed parameter
space that generates the relic density through
stau-neutralino coannihilation. Hence, the five lightest sparticles for this benchmark point are
$\widetilde{\chi}_{1}^{0} < \widetilde{\tau}_{1}^{\pm} < \widetilde{e}_{R} < \widetilde{\chi}_{2}^{0}
\sim \widetilde{\chi}_{1}^{\pm}$. Here, the gluino is lighter than all the squarks with the exception
of the lightest stop, so all squarks will predominantly decay to a gluino and hadronic jet,
with a small percentage of squarks producing a jet and either a $\widetilde{\chi}_{1}^{\pm}$ or
$\widetilde{\chi}_{2}^{0}$. The gluinos will decay via virtual (off-shell) squarks to neutralinos or charginos
plus quarks, which will further cascade in their decay. The result is a
low-energy tau through the processes $\widetilde{\chi}_{2}^{0} \rightarrow \widetilde{\tau}_{1}^{\mp}
\tau^{\pm} \rightarrow \tau^{\mp}\tau^{\pm} \widetilde{\chi}_{1}^{0}$ and $\widetilde{\chi}_{1}^{\pm}
\rightarrow \widetilde{\tau}_{1}^{\pm} \nu_{\tau} \rightarrow \tau^{\pm}\nu_{\tau} \widetilde{\chi}_{1}^{0}$.

The LHC final states of low-energy tau in the ${\cal F}$-$SU(5)$ stau-neutralino coannihilation region
are similar to those same low-energy LHC final states in mSUGRA, however, in the stau-neutralino
coannihilation region of mSUGRA, the gluino is typically heavier than the squarks. The strong coupling
effects from the additional vector-like particles on the gaugino mass RGE running reduce the physical
gluino mass below the squark masses in ${\cal F}$-$SU(5)$. As a consequence, the LHC final low-energy
tau states in the stau-neutralino coannihilation regions of ${\cal F}$-$SU(5)$ and mSUGRA will differ
in that in ${\cal F}$-$SU(5)$, the low-energy tau states will result largely from neutralinos and
charginos produced by gluinos, as opposed to the low-energy tau states in mSUGRA resulting primarily
from neutralinos and charginos produced from squarks.

Also notably, the TeV-scale vector-like multiplets are well targeted
for observation by the LHC.  We have argued~\cite{Li:2010mi} that the eminently
feasible near-term detectability of these hypothetical fields in collider experiments,
coupled with the distinctive flipped charge assignments within the multiplet structure,
represents a smoking gun signature for Flipped $SU(5)$, and have thus coined the term
{\it flippons} to collectively describe them.
Immediately, our curiosity is piqued by the recent announcement~\cite{Abazov:2010ku}
of the D\O~collaboration that vector-like quarks have been excluded up to
a bound of 693~GeV, corresponding to the immediate lower edge of our anticipated
range for their discovery~\cite{Li:2010mi}.

%%%%%%%%%%%%%%%%%%%%%%%%%%%%%%%%%%%%%%%%%%%%%%%%%%%%%%%%%%%%%%%%%%%%%%%%%%%%%

\section{Conclusion}

The advancement of human scientific knowledge and technology is replete with instances of science fiction transitioning to scientific
theory and eventually scientific fact. The conceptual notion of a ``Multiverse'' has long fascinated the human imagination,
though this speculation has been largely devoid of a substantive underpinning in physical theory.  The modern perspective presented
here offers a tangible foundation upon which legitimate discussion and theoretical advancement of the Multiverse may commence,
including the prescription of specific experimental tests which could either falsify or enhance the viability of our proposal.
Our perspective diverges from the common appeals to statistics and the anthropic principle, suggesting instead that we may seek
to establish the character of the master theory, of which our Universe is an isolated vacuum condensation, based on specific observed properties
of our own physics which might be reasonably inferred to represent invariant common characteristics of all possible universes.  We have
focused on the discovery of a model universe consonant with our observable phenomenology, presenting it as confirmation of a
non-zero probability of our own Universe transpiring within the larger String Landscape.

The archetype model universe which we advance in this work implicates No-Scale Supergravity as
the ubiquitous supporting structure which pervades the vacua of the Multiverse,
being the crucial ingredient in the emanation of a cosmologically flat universe from the quantum ``nothingness''.
In particular, the model dubbed No-Scale ${\cal F}$-$SU(5)$ has demonstrated remarkable consistency between parameters
determined dynamically (the top-down approach) and parameters determined through the application of current
experimental constraints (the bottom-up approach).  This enticing convergence of theory with experiment
elevates No-Scale ${\cal F}$-$SU(5)$, in our estimation, to a position as the current leading GUT candidate.
The longer term viability of this suggestion is likely to be greatly clarified in the next few years, based upon the wealth of
forthcoming experimental data.

Building on the results presented in prior works~\cite{Li:2010ws,Li:2010mi,Li:2010uu}, we have presented a dynamic determination
of the penultimate Flipped $SU(5)$ unification scale $M_{32}$, or more fundamentally, the GUT Higgs VEV moduli. We have demonstrated
that the $B_{\mu}$ = 0 No-Scale boundary condition is again vital in dynamically determining the model parameters.
Procedurally, we have fixed the unified gauge coupling, SM fermion Yukawa couplings, and Higgs bilinear term $\mu\simeq 460~{\rm GeV}$
at the final unification scale $M_{\cal F}$, while concurrently allowing the VEVs of the GUT Higgs fields $H$ and $\overline{H}$ to float freely,
as driven by $M_{32}$ and the low energy SM gauge couplings, via variation of the Weinberg angle.  Employing the ``Super No-Scale''
condition to secondarily minimize the effective Higgs potential, we have obtained $M_{32}\simeq 1.0 \times 10^{16}$~GeV,
$\sin^2 (\theta_{\rm W}) \simeq 0.236$, and $\tan \beta \simeq 15-20$.

The blueprints which we have outlined here, integrating precision phenomenology with prevailing experimental data and
a fresh interpretation of the Multiverse and the Landscape of String vacua, offer a
logically connected point of view from which additional investigation may be mounted.
As we anticipate the impending stream of new experimental data which is likely to be revealed in ensuing years,
we look forward to serious discussion and investigation of the perspective presented in this work.
Though the mind boggles to contemplate the implications of this speculation, so it must also
reel at even the undisputed realities of the Universe, these acknowledged facts alone being
manifestly sufficient to humble our provincial notions of longevity, extent, and largess.

%%%%%%%%%%%%%%%%%%%%%%%%%%%%%%%%%%%%%%%%%%%%%%%%%%%%%%%%%%%%%%%%%%%%%%%%%%%%%

\section*{Acknowledgments}

This research was supported in part 
by  the DOE grant DE-FG03-95-Er-40917 (TL and DVN),
by the Natural Science Foundation of China 
under grant No. 10821504 (TL),
and by the Mitchell-Heep Chair in High Energy Physics (TL).

%%%%%%%%%%%%%%%%%%%%%%%%%%%%%%%%%%%%%%%%%%%%%%%%%%%%%%%%%%%%%%%%%%%%%%%%%%%%%

\end{document}